\documentclass[twocolumn,nofootinbib,showpacs,amsmath,amssymb,aps,prd,balancelastpage,ctexart,superscriptaddress]{revtex4-1}

\usepackage{epsfig}
\usepackage{bm}
\usepackage{verbatim}
\usepackage[latin1]{inputenc}
\usepackage{color}
\usepackage{float}
\usepackage{tikz}
\usepackage{dcolumn}

\usepackage{amsmath,amssymb}
\usepackage{amsfonts,amssymb,mathrsfs}
\usepackage{graphicx}
\usepackage[bookmarks=false,pdfstartview=FitH]{hyperref}
\usepackage[all]{hypcap}

\def\ba{\begin{align}}
\def\ea{\end{align}}
\def\be{\begin{equation}}
\def\ee{\end{equation}}
\def\bea{\begin{eqnarray}}
\def\eea{\end{eqnarray}}

\begin{document}

\title{Superconducting cosmic strings as sources of cosmological fast radio bursts}

\author{Jiani Ye}
\affiliation{CAS Key Laboratory for Research in Galaxies and Cosmology, Department of Astronomy, University of Science and Technology of China, Hefei, Anhui 230026, China}
\affiliation{Shanghai Astronomical Observatory, Chinese Academy of Sciences, Shanghai 200030, China}
\affiliation{Department of Physics and Astronomy, Stony Brook University, Stony Brook, New York 11794, USA}
\affiliation{University of Chinese Academy of Sciences, Beijing, China}

\author{Kai Wang}
\affiliation{CAS Key Laboratory for Research in Galaxies and Cosmology, Department of Astronomy, University of Science and Technology of China, Hefei, Anhui 230026, China}
\affiliation{School of Astronomy and Space Science, University of Science and Technology of China, Hefei, Anhui 230026, China}

\author{Yi-Fu Cai}
\email{yifucai@ustc.edu.cn}
\affiliation{CAS Key Laboratory for Research in Galaxies and Cosmology, Department of Astronomy, University of Science and Technology of China, Hefei, Anhui 230026, China}
\affiliation{School of Astronomy and Space Science, University of Science and Technology of China, Hefei, Anhui 230026, China}

\begin{abstract}
In this paper we calculate the radio burst signals from three kinds of structures of superconducting cosmic strings.
By taking into account the observational factors including scattering and relativistic effects, we derive the event rate of radio bursts as a function of redshift with the theoretical parameters $G\mu$ and $\mathcal{I}$ of superconducting strings. Our analyses show that cusps and kinks may have noticeable contributions to the event rate and in most cases cusps would dominate the contribution, while the kink-kink collisions tend to have secondary effects. By fitting theoretical predictions with the normalized data of fast radio bursts, we for the first time constrain the parameter space of superconducting strings and report that the parameter space of $G\mu \sim [10^{-14}, 10^{-12}]$ and $\mathcal{I} \sim [10^{-1}, 10^{2}] ~ \rm{GeV}$ fit the observation well although the statistic significance is low due to the lack of observational data. Moreover, we derive two types of best fittings, with one being dominated by cusps with a redshift $z = 1.3$, and the other dominated by kinks at the range of the maximal event rate.
\end{abstract}

\pacs{98.80.Cq, 11.27.+d, 95.85.Bh, 95.85.Fm}

\maketitle

\section{Introduction}\label{Introduction}

Cosmic strings can be formed in phase transitions of the very early universe if the manifold of the background vacuum has nontrivial topology for the spontaneous symmetry breaking (see Refs. \cite{Kibble:1976sj, Vilenkin:2000jqa, Polchinski:2004ia, Copeland:2009ga, Ringeval:2010ca, Copeland:2011dx, Vachaspati:2015cma} for comprehensive reviews). The potential exploration of these primordial relics could reveal important information as regards fundamental physics at extremely high energy scales, in which one class of particle physics models naturally give rise to cosmic strings that are superconducting \cite{Witten:1984eb}. Superconducting cosmic strings (SCSs), as can be realized by a charged scalar field whose flux is trapped in normal cosmic strings with the electromagnetic gauge invariance broken inside the strings, can produce electromagnetic effects \cite{Vilenkin:2000jqa}. These superconductive wire-like objects in the sky behave as giant antennas that can emit electromagnetic signals in a wide range of frequencies \cite{Vilenkin:1986zz, Garfinkle:1987yw}.

Due to the oscillations of cosmic strings, they are able to emit electromagnetic radiation as well as gravitational radiation and thus can give rise to various observational effects of SCSs. These effects have been extensively studied in the literature, such as the distortions and anisotropies in the cosmic microwave background (CMB) caused by electromagnetic or gravitational radiations of cosmic strings \cite{Sanchez:1988ek, Sanchez:1990kj, Tashiro:2012nb, Jazayeri:2017szw}, the early reionization by SCSs and the associated imprints in the CMB \cite{Tashiro:2012nv}, the high energy cosmic rays due to the decay of cosmic strings \cite{Hill:1986mn, Bhattacharjee:1989vu, Bhattacharjee:1990js, MacGibbon:1989kk, Wichoski:1998kh, Brandenberger:2009ia, Berezinsky:2009xf, Lunardini:2012ct}, the gamma-ray bursts induced by cosmic strings \cite{Babul:1987lza, Paczynski:1988, Brandenberger:1993hw, Plaga:1993hp, Berezinsky:2001cp, Cheng:2010ae}, gravitational waves \cite{Vachaspati:1984gt, Damour:2000wa, Damour:2001bk, Damour:2004kw}, as well as sparks \cite{Vachaspati:2008su} and radio transients from SCSs \cite{Cai:2011bi, Cai:2012zd, Yu:2014gea}. Accordingly, the parameter space of SCSs, which is often characterized by the string tension $\mu$ (or $G\mu$ in Planck units) and the current on the string $\mathcal{I}$, can be derived from the comparison with various observational data. For instance, the CMB analysis based on WMAP and the South Pole Telescope yields an upper bound on the string tension as $G\mu < 1.7 \times 10^{-7}$ \cite{Dvorkin:2011aj}, and this constraint has been recently improved to be $G\mu < 1.3 \times 10^{-7}$ with the help of the Planck data \cite{Ade:2013xla}. As cosmic strings can also lead to relic gravitational waves in the form of a stochastic background, a model-dependent constraint from the pulsar timing measurements on the string tension can be obtained in about the same order of the bound from CMB \cite{vanHaasteren:2011ni, Pshirkov:2009vb, Tuntsov:2010fu, Olmez:2010bi, Binetruy:2012ze, Sanidas:2012ee}. Further constraints on SCSs can be derived through the spectral distortions of the CMB photons \cite{Sanchez:1988ek, Sanchez:1990kj, Tashiro:2012nb}, namely, the parameter region with $10^{-19} < G\mu < 10^{-7}$ and $\mathcal{I} > 10^4 ~ {\rm GeV}$ would be ruled out \cite{Kogut:2011xw}.

The signals of electromagnetic radiations from SCSs can be significantly enhanced if they were emitted from particular structures of strings, such as the cusps \cite{Vachaspati:2008su}, and the kinks \citep{Cai:2011bi, Cai:2012zd}. These signals are found to be in form of radio bursts, and thus, the SCSs could serve as the possible sources of the Fast Radio Burst (FRB) events that have been discovered in astronomical observations. So far, there have been $18$ FRB events reported by various radio experiments including the Parkes \cite{Lorimer:2007qn, Keane:2011mj, Thornton:2013iua, Burke-Spolaor:2014rqa, Petroff:2014taa, Ravi:2014mma, Champion:2015pmj, Keane:2016yyk, Ravi:2016kfj}, the Arecibo Pulsar ALFA Survey \cite{Spitler:2014fla}, and the Green Bank Telescope \cite{Masui:2015kmb}. The events of FRB are characterized as having a bandwidth of the millisecond scale, and frequency around ${\rm GHz}$ scale. All these events show extra-galactic origins and their estimated redshifts are within $z=1.4$. The flux of these events are about several ${\rm Jys}$. Theoretically there have been some attempts to provide the sources of these FRB signals, such as super-massive neutron stars \cite{Falcke:2013xpa}, binary neutron stars or white dwarf mergers \cite{Totani:2013lia, Kashiyama:2013gza}, local circumnuclear magnetars \cite{Pen:2015ema}, and so on.

In the present study we investigate the radio burst signals from SCSs using the method of order-of-magnitude analysis. We compare the normalized FRB event rate as a function of redshift based on theoretical models with normalized observational data, and thus derive the limits on theoretical parameters of strings. Although there are not enough data to offer sufficiently precise results, our analysis provides a theoretically sound method of constraining theoretical parameters of SCSs in the ongoing search for FRBs at the Parkes \cite{Lorimer:2007qn}, ETA \cite{Patterson:2008ie}, LWA \cite{Henning:2010aa}, LOFAR \cite{Fender:2008zh} , and FAST \cite{Nan:2011um} radio telescopes. We note that an attempt on the astronomical limit of SCSs based on the specific data of a single FRB event detected by the Parkes was investigated in \cite{Miyamoto:2012ck}, and later, the work of \cite{Yu:2014gea} has discussed FRBs as a possible probe on model parameters of SCSs such as string length and $\mu$ by ascribing four FRB events. The model of calculating the radiations from SCSs applied in these two works only focused on the particular structure of string cusps. However, as we shall show in the present work, radiations from kinks can also lead to significant contribution under certain conditions. Moreover, we in the present study take into account the scattering effects from the interstellar medium (ISM) and intergalactic medium (IGM) as well as the statistical distributions of string loops. With the help of accumulated astronomical data, our analysis shows that the statistical bias can be greatly reduced and hence the astronomical constraints on the parameter space of SCSs from the FRB events would be statistically meaningful in near future.

The structure of the present paper is as follows. In Sec.~\ref{sec:theo} we describe the general characteristics of SCSs by reviewing the radiation emission from these charged wires in the sky. Then in Sec.~\ref{sec:rad} we perform an order-of-magnitude analysis method to solve out the equations of electromagnetic power emitted from string structures including cusps and kinks. Sec.~\ref{sec:eventrate} is devoted to the calculation of the theoretical prediction on the event rate of radio bursts as a function of the redshift, flux and string length parameters. Afterwards, we in Sec.~\ref{sec:numer&data} perform the numerical estimation in detail by comparing the theoretical prediction with the observational data of FRBs, and then report a brand-new constraint on the parameter space of cosmic strings. We summary the results with a discussion in Sec~\ref{sec:discussion}. In the theoretical calculations we use natural units, $\hbar = c = 1$.

\section{Characteristics of SCSs} \label{sec:theo}

To begin with, we consider the effective action of a superconducting string as follows \cite{Vilenkin:2000jqa}:
\begin{align}
 S = \int d^2\zeta & \bigg\{ -\mu \sqrt{-\gamma} +\frac{1}{2} \sqrt{-\gamma} \gamma^{ab} \phi_{,a}\phi_{,b} -A_{\mu} x^{\mu}_{,a} J^{a} \bigg\} \nonumber\\
 & - \frac{1}{16\pi} \int d^4x \sqrt{-g} F_{\mu\nu} F^{\mu\nu} ~,
\end{align}
where $\gamma_{ab}$ is the induced metric on the worldsheet of a string with $(a, b) = 0\ {\rm or}\ 1$, and $\mu$ is the string tension. The first term in the rhs of the above action is the Nambu-Goto action; the second and third terms represent for the worldsheet current and the coupling to the electromagnetic gauge potential, respectively, with $x^\mu$ being introduced as the string position, $\phi$ the massless real scalar field on the worldsheet, and $A_{\mu}$ the gauge field of the worldsheet current $J^{a}$; and the last term describes the dynamics of electromagnetic field in the four dimensional spacetime.

The current on the worldsheet is given by
\begin{equation}
 J^{a} = q\epsilon^{ab}\phi_{,b}/\sqrt{-\gamma} ~,
\end{equation}
where $q$ is the charge of the current carriers. We focus on the classical production of bursts and at the moment simplify the metric of the background spacetime as the Minkowski $\eta_{\mu\nu}$. We choose the conformal gauge for the worldsheet, which can be specified by the following conditions:
\begin{equation}
\label{eq:conformal}
 \dot{x}^{\mu}x'_{\mu}=0 ~,~~ \dot{x}^2+x'^{2}=0 ~,
\end{equation}
where ``~$\dot{~} \equiv \partial/\partial\tau$'' and ``~$' \equiv \partial/\partial\sigma$'' has been introduced, and $(\tau, \sigma)$ correspond to the worldsheet coordinates. Under this condition, the induced worldsheet metric is then simplified as $\gamma^{ab}=\rm{diag}(1,-1)$. Consequently, we write down the equations of motion for the scalar field, the string and the gauge field respectively as,
\begin{align}
\label{eq:phi}
 & \Box_2 \phi = -\frac{1}{2}q \epsilon^{ab} F_{\mu\nu} x^{\mu}_{,a} x^{\nu}_{,b} ~, \\
\label{eq:x}
 & \Box_2 x^{\mu} = \big[ (-F^{\mu}_{~\sigma}x^\sigma_{,a} J^a - (\Theta^{ab}x^\mu_{,a})_{,b} \big] / \mu ~, \\
\label{eq:A}
 & \Box_4 A^{\mu} = 4\pi j^{\mu} ~.
\end{align}
In the above equations of motion, we have introduced
\begin{align}
\label{eq:box}
 & \Box_2 = \eta^{ab} \partial_a \partial_b ~,~ \Box_4 = \eta^{\mu\nu}\partial_{\mu}\partial_{\nu} ~, \\
\label{eq:Thetaab}
 & \Theta_{ab} = \phi_{,a}\phi_{,b} - \frac{1}{2} \gamma_{ab}\phi_{,c}\phi^{,c} ~, \\
\label{eq:current}
 & j^{\mu}(x) = \int d\sigma d\tau x^{\mu}_{,a} \sqrt{-\gamma} ~ J^{a} ~ \delta^{(4)}\big( x-x(\sigma, \tau) \big) ~,
\end{align}
where $\Theta_{ab}$ is the stress energy tensor of the scalar field living on the worldsheet and $j^\mu$ corresponds to the current in the four dimensional spacetime. We adopt the Lorentz gauge for $A^{\mu}$: $\partial_{\mu}A^{\mu} = 0$. In order to fix another degree of freedom, we also set $x^0=\tau=t$ with $t$ being the physical time. We use $t$ instead of $\tau$ to denote $x^0$ in the following discussions unless otherwise claimed.

In the center-of-mass frame of a string loop with an invariant length $L$, we can calculate the solution for Eq. \eqref{eq:x} to be
\begin{equation}
\label{eq:x_pm}
 x^{\mu}(t, \sigma) = \frac{1}{2} \big[ x^{\mu}_{-}(\sigma_{-}) + x^{\mu}_{+}(\sigma_{+}) \big] ~,
\end{equation}
where $\sigma_{\pm} \equiv \sigma \pm t$. Since the periodicity of $\sigma$ takes the string length $L$, from the solution we can deduce that the periodicity of $t$ is $L/2$. To combine the gauge conditions in Eq. \eqref{eq:conformal} with the solution, we can get
\begin{equation}
\label{eq:gaugeconditions}
 x^0_{-} = -\sigma_{-},x^0_{+} = \sigma_{+} ~,~ |\mathbf{x}'_{+}| = 1 = |\mathbf{x}'_{-}| ~.
\end{equation}
As a result, for a string with length $L$, the order of magnitude of $\mathbf{x}''_{\pm}$ and $\mathbf{x}'''_{\pm}$ can be estimated to be ${1}/{L}$, ${1}/{L^2}$ respectively.

Note that the vanishing rhs of Eq. \eqref{eq:phi} implies a conservation of the current $J^a$ on the worldsheet. To integrate Eq. \eqref{eq:current} over $t$ for a period, we then have
\begin{equation}
\label{eq:j-2}
 j^{\mu}(t, \mathbf{x}) = \mathcal{I} \int d\sigma x^{\mu}_{,\sigma} \delta^{(3)}\big( \mathbf{x} -\mathbf{x}(t,\sigma) \big) ~,
\end{equation}
where $\mathcal{I}$ is the parameter of the current on the string. After that, we perform the Fourier transform
\begin{align}
\label{eq:j-3}
 j^{\mu}(t, \mathbf{x} ) = \underset{\omega}{\Sigma} \int \frac{d^3 k}{(2\pi)^3} e^{-i (\omega t - \mathbf{k} \cdot \mathbf{x})} j^{\mu}_{\omega}(\mathbf{k}) ~,
\end{align}
which yields
\begin{align}
 j^{\mu}_{\omega}(\mathbf{k}) = \frac{2\mathcal{I}}{L} \int^{L/2}_{0} dt \int^{L}_{0} d\sigma e^{ik_{\mu}x^{\mu}} x^{'\mu} \equiv \frac{\mathcal{I}}{L} (I^{\mu}_{n+}+I^{\mu}_{n-}) ~.
\end{align}
In the above equation we have applied the definition of $x^{\mu}_{\pm}$ and thus there is
\begin{equation}
 I^{\mu}_{n\pm} = \int^{L/2}_{0} dt \int^{L}_{0} d\sigma_{\pm} e^{ik_{\mu}x^{\mu}_{\pm}} x^{'\mu}_{\pm} ~.
\end{equation}
The electromagnetic radiation power from a periodic source is then given by
\begin{align}
\label{eq:power}
 P &= \underset{n}{\Sigma} P_{\omega_n} ~, \\
\label{eq:powern}
 P_{\omega_n} &= -\frac{L}{2\pi} \omega_n^2 \int d\Omega ~ j^{*}_{\mu}(\omega_n, \mathbf{k}) ~ j^{\mu}(\omega_n, \mathbf{k}) ~,
\end{align}
where $\omega_n$ denotes the Fourier mode with $\omega_n = 4\pi n/L$, $n\in \mathbf{N}$. From the expression of $P_{\omega_n}$, we can see that the key to estimating the radiation power from SCSs is to solve out the solution of $j^{\mu}_{\omega}$. The main approach of releasing electromagnetic radiation from the string loop is to study the emission processes from certain structures including cusps, kinks and kink-kink collisions. Other than these structures, the radiation power would decrease exponentially and hence is ignored for simplicity. In the study of gravitational or electromagnetic radiations from cosmic strings, it is common to make use of the saddle point/discontinuity approximation method to find the corresponding solutions (for instance see Refs. \citep{Damour:2004kw, Steer:2010jk, Cai:2012zd}). In order to carry out more details of the same processes, we in the present analysis follow the method in Refs. \citep{Vilenkin:1986zz, BlancoPillado:2000xy}, which was adopted in investigating the string cusps, and, perform a much illustrative and generalized approach to solve the solutions to particular structures of cosmic strings.

\section{Radiations from SCSs} \label{sec:rad}

The radiations from string loops are mostly contributed by the areas where $j^{\mu}_{\omega}$ takes the maximal value. As we will show later, outside these areas the contributions would decrease exponentially. For convenience, one can set the position of the maximal value for $j^{\mu}_{\omega}$ to be the origin point $(t,\sigma) = (0, 0)$, and then express $j^{\mu}_{\omega}$ in the following form:
\begin{align}
\label{eq:j-4}
 j^{\mu}_{\omega}(\mathbf{k}) &= \frac{2\mathcal{I}}{L} \int^{L/2}_{0} dt \int^{L}_{0} d\sigma ~ e^{i(\omega t -\mathbf{k} \cdot \mathbf{x})} ~ x^{'\mu} \nonumber\\
 &= \frac{2\mathcal{I}}{L} \int^{L/2}_{0} dt \int^{L}_{0} d\sigma ~ e^{i(\omega t -\mathbf{k} \cdot \mathbf{x}_{+}/2 -\mathbf{k} \cdot \mathbf{x}_{-}/2)}
 \frac{1}{2}\big[ x^{'\mu}_{+} +x^{'\mu}_{-} \big] ~.
\end{align}
The wave vector $\mathbf{k}$ is also identified as $\mathbf{k}=\omega\mathbf{n}$. Note that Eq. \eqref{eq:gaugeconditions} yields $j^0_{\omega}=0$. Therefore, in the following study we focus on the spatial part of the four current $j^{\mu}_{\omega}$: $\mathbf{J}_{\omega}$. The maximal value position requires $(\omega t - \mathbf{k}\cdot\mathbf{x})' = 0$, where $(\omega t - \mathbf{k}\cdot\mathbf{x})$ is the exponential factor in the expression of $j^{\mu}_{\omega}$. From this requirement, one derives
\begin{equation}
\label{eq:cuspconditions}
 \mathbf{n} \cdot \mathbf{x}_{+}' +\mathbf{n} \cdot \mathbf{x}_{-}' =0 ~,
\end{equation}
and for $\mathbf{n}$ not perpendicular to $\mathbf{x}_{\pm}'$, it is equivalent to that $\mathbf{x}_{+}'+\mathbf{x}_{-}'=0$. Such particular positions are denoted as $\it{cusps}$, if both
$\mathbf{x}_{+}'$ and $\mathbf{x}_{-}'$ are continuous; or, $\it{kinks}$, if one of $\mathbf{x}_{\pm}'$ is discontinuous; or even, $\it{kink-kink\ collisions}$, if both $\mathbf{x}_{\pm}'$ are discontinuous. We analyze the electromagnetic radiation power emitted around these particular structures, respectively.

\subsection{Cusps} \label{cusp}


In the neighborhood of the origin point $(t,\sigma)=(0,0)$, we can perform the Taylor expansions of $\mathbf{x}_{\pm}$ as follows:
\begin{align}
 \mathbf{x}_{+} &= \mathbf{x}_{+0}{'}(\sigma+t) +\frac{1}{2}\mathbf{x}_{+0}{''}(\sigma+t)^2 +\frac{1}{6}  \mathbf{x}_{+0}{'''}(\sigma+t)^3 + \cdots, \\
 \mathbf{x}_{-} &= \mathbf{x}_{-0}{'}(\sigma-t) +\frac{1}{2}\mathbf{x}_{-0}{''}(\sigma-t)^2 +\frac{1}{6}  \mathbf{x}_{-0}{'''}(\sigma-t)^3 + \cdots ~,
\end{align}
and these expressions are valid as long as the scales involved are smaller than the string length $\mathcal{O}(L)$).
To substitute Eq. \eqref{eq:gaugeconditions} into the above expansions, one obtains
\begin{align}
\label{eq:xcond1}
 & \mathbf{x}_{+0}{'} = -\mathbf{x}_{-0}{'} ~, \\
\label{eq:xcond2}
 & \mathbf{x}_{+0}{'} \cdot \mathbf{x}_{+0}{''} = \mathbf{x}_{-0}{'} \cdot \mathbf{x}_{-0}{''} = 0 ~,
\end{align}
and accordingly, from Eq. \eqref{eq:j-4} the form of $\mathbf{J}_{\omega}$ can be estimated to be
\begin{align}
\label{eq:jmuk_cusp}
 \mathbf{J}_{\omega}(\mathbf{k}) & \simeq \frac{2\mathcal{I}}{L} |t_*| \int _0^L \frac{d\sigma}{2} \Big[ {\mathbf{x}_{+0}' +\mathbf{x}_{-0}'} +\mathbf{x}_{+0}''(\sigma+t) +\mathbf{x}_{-0}''(\sigma-t) +\cdot\cdot\cdot \Big] \nonumber\\
 & \sim \frac{2\mathcal{I}}{L}| t_{*}|\frac{1}{L}|\sigma_*|^2 ~.
\end{align}

In the above formulae, $t_*$ and $\sigma_*$ represents for the characteristic regime when the radiations dominate over. In order to estimate their magnitudes, we apply the condition $(\omega t-\mathbf{k}\cdot \mathbf{x})\simeq 1$ and then perform the Taylor expansion of $\mathbf{k}\cdot \mathbf{x}$ around $(t,\sigma)=(0,0)$ as follows:
\begin{align}
\label{eq:wt-kx}
 \omega t - \mathbf{k}\cdot \mathbf{x} 
 = & \omega \Big\{ t-\frac{1}{2} \mathbf{n} \cdot\big[ \mathbf{x}_{+0}'(\sigma+t) + \mathbf{x}_{-0}'(\sigma-t) \big] \nonumber\\
 & - \frac{1}{4} \mathbf{n} \cdot \big[ \mathbf{x}_{+0}''(\sigma+t)^2
 +\mathbf{x}_{-0}''(\sigma-t)^2 \big] \nonumber\\
 & - \frac{1}{12} \mathbf{n}\cdot \big[\mathbf{x}_{+0}'''(\sigma+t)^3 +\mathbf{x}_{-0}'''(\sigma-t)^3\big] +\cdot\cdot\cdot \Big\} ~.
\end{align}
Notice that approximately $\mathbf{x}_{\pm}'' \sim 1/L$ and $\mathbf{x}_{\pm}''' \sim 1/L^2$.
Moreover, Eq. \eqref{eq:xcond1} shows that $\mathbf{x}_{+}'$ and $\mathbf{x}_{-}'$ are anti-parallel, and thus, from Eq. \eqref{eq:cuspconditions} one can learn that the orders of magnitude for $(t, \sigma)$ should be small quantities. In addition, if $\mathbf{n}$ is parallel to either $\mathbf{x}_{+}'$ or $\mathbf{x}_{-}'$, Eqs. \eqref{eq:gaugeconditions}, \eqref{eq:cuspconditions} and \eqref{eq:xcond2} then ensure $\mathcal{O}(t^2, \sigma^2)$ to be also negligible.
As a consequence, we can reach $\omega \mathcal{O}(t^3, \sigma^3)/L^2\sim 1$, which implies that the most possible orders of magnitude for $|t_*|$ and $|\sigma_*|$ in the case of cusps can be estimated as
\begin{equation}
\label{eq:tsigmacusp}
 |t_*| ,|\sigma_*| \sim L^{2/3}/\omega^{1/3}.
\end{equation}

Similar to the above process, we can also estimate the order of magnitude of the solid angle $\Omega$ as follows. We consider that $\mathbf{n}$ has a small deviation from $\mathbf{x}_{\pm}'$ within the range of the angle $\theta$. Without loss of generality, we take $\mathbf{n}\cdot\mathbf{x}_{+}' = cos \theta$, $\mathbf{n}\cdot\mathbf{x}_{-}' = cos (\pi-\theta)$, and hence, from Eq. \eqref{eq:wt-kx} get the following approximations:
\begin{equation}
\label{eq:thetalim}
 \omega t - \mathbf{k}\cdot \mathbf{x} \sim {\theta^2}{\omega t} \sim \theta \omega L^{-1}(t^2,t\sigma, \sigma^2) \sim 1 ~,
\end{equation}
which further yields
\begin{equation}
 \theta \sim \frac{1}{(\omega L)^{1/3}} ~.
\end{equation}

All possible deviations of the angle $\theta$ can form a cone-shape region around $\mathbf{x}_{\pm}'$, and thus, the solid angle scope where the radiations could be observed from a cusp is limited within
\begin{equation}
\label{eq:Thetacusp}
 \Omega \sim \theta^2 \sim \frac{1}{(\omega L)^{2/3}} ~.
\end{equation}
Moreover, the radiations from the cusp would decrease exponentially outside the solid angle $\Omega$, which implies that the electromagnetic signals emitted from a cusp are expected to be highly focused beams. On substituting Eq. \eqref{eq:tsigmacusp} into Eq. \eqref{eq:jmuk_cusp}, we get $\mathbf{J}_{\omega} \sim \frac{\mathcal{I}}{\omega}$ approximately. Using Eq. \eqref{eq:powern}, the power emitted by string cusps at the frequency $\omega$ can be estimated as
\begin{equation}
\label{eq:cuspp}
 P_{\omega}=\frac{\mathcal{I}^2 L^{1/3}}{\omega^{2/3}} ~.
\end{equation}
The total power emitted from this type of the string structure can be obtained by integrating $P_{\omega}$ over the frequency $\omega$. This result looks divergent. This is because that cosmic strings actually remains the inner structure characterized by its tension $\mu$. Therefore, we can provide an upper limit of $\omega$ by the theoretical constraint that, given the tension $\mu$ of a string, the maximum energy of radiations emitted from this string ought to be less than the energy of the string itself. As a result, the radiation power from cusps can then be estimated as~\cite{Vilenkin:1986zz, Cai:2012zd}
\begin{equation}
P^{c}_{\gamma} \sim \mathcal{I}^2(\omega_{max}L)^{1/3}\sim \kappa\mathcal{I}\sqrt{\mu}.
\label{eq:cusplimitL}
\end{equation}
Numerical simulations yield a result in the same order of magnitude, with a correction of the factor $\kappa\sim 10$.

\subsection{Kinks}\label{kink}

In the case of string kinks, either ${x}^{\mu}_{+}{'}$ or ${x}^{\mu}_{-}{'}$ is discontinuous. Without loss of generality, we set that ${x}^{\mu}_{+}{'}$ is discontinuous at $\sigma =0 $. Similar to Eq. \eqref{eq:wt-kx}, we make the Taylor expansion of $(\omega t-\mathbf{k}\cdot\mathbf{x})$ from both sides, which is expressed as
\begin{align}
\label{eq:kink-we-kx}
 (\omega t - \mathbf{k}\cdot \mathbf{x})_{(0\pm)} &\sim \omega \Big\{ t - \frac{1}{2} \big[ \mathbf{n}\cdot\mathbf{x}_{+}{'}_{(0\pm)}(\sigma+t) +\mathbf{n}\cdot\mathbf{x}_{-}'(\sigma-t) \big] \nonumber\\
 & + \frac{1}{4} \big[ \mathbf{n}\cdot\mathbf{x}_{+}{''}_{(0\pm)}(\sigma+t)^2 +\mathbf{n}\cdot\mathbf{x}_{-}''(\sigma-t)^2 \big] \nonumber\\
 & + \frac{1}{12} \big[ \mathbf{n}\cdot\mathbf{x}_{+}{'''}_{(0\pm)}(\sigma+t)^3 +\mathbf{n}\cdot\mathbf{x}_{-}'''(\sigma-t)^3 \big] \nonumber\\
 & +\cdot\cdot\cdot \Big\} ~,
\end{align}
where the subscript $(0\pm)$ denotes the expansions from the positive and negative side, respectively. The coefficient $\sigma$ takes $-\frac{1}{2}(\mathbf{n}\cdot\mathbf{x}_{+}{'}_{(0\pm)}+\mathbf{n}\cdot\mathbf{x}_{-}{'})$, where $\mathbf{x}_{-}{'}$ is parallel with one side of $\mathbf{x}_{+}{'}$ by assuming that it is $\mathbf{x}_{+}{'}_{(0+)}$.
From Eq. \eqref{eq:cuspconditions}, this coefficient must vanish, thus the direction of $\mathbf{n}$ can only be in a plane such that the angle between $\mathbf{n}$ and $\mathbf{x}_{+}{'}_{(0+)}$ equals that between $\mathbf{n}$ and $\mathbf{x}_{+}{'}_{(0-)}$. We can then check the $\sigma^2$ coefficient. Note that, for both sides of the kinks, Eqs. \eqref{eq:cuspconditions} and \eqref{eq:xcond2} remain applicable, but here we can only derive $\mathbf{n}~\cdot~\mathbf{x}_{+}{''}_{(0\pm)}~+~\mathbf{n}~\cdot~\mathbf{x}_{-}{''}~=~0$, and accordingly, the $\sigma^2$ coefficient requires $\sigma \sim L/(\omega |t_*|)$; however, the $\sigma^3$ coefficient further requires $\sigma \sim L^{2/3}/\omega^{1/3}$. To take into account both estimations, one can set the final form of $|\sigma_*|$ to be the minimal of these two results. Given the relations between $\mathbf{n}$ and $\mathbf{x}_{+}{'}_{(0\pm)},\mathbf{x}_{-}{'}$, the first-order coefficient of $t$ would not vanish. As a result, we have $|t_*|\sim \omega^{-1}$. Substituting $|t_*|$ into the $\sigma^2$ coefficient constraint, we find that the dominant order of magnitude for $\sigma$ comes from the $\sigma^3$ term in the case of kinks. To conclude, $(\omega t-\mathbf{k} \cdot \mathbf{x})\sim 1$ yields the order of $t$ and $\sigma$ for kinks as follows:
\begin{equation}
\label{eq:tsigmakink}
 |t_{*}|\sim \omega^{-1},\ |\sigma_{*}|\sim L^{2/3}/\omega^{1/3} ~.
\end{equation}

Unlike cusps, for kinks, the constraint of the angle of the radiation emission, as we have discussed above, has only one dimension. By substituting Eq. \eqref{eq:tsigmakink} into \eqref{eq:thetalim}, one gets
\begin{equation}
\label{eq:Thetakink}
 \Omega \sim \theta \sim \frac{1}{(\omega L)^{1/3}} ~,
\end{equation}
We note that the only restriction of $\theta$ comes from the requirement of $\theta\omega L^{-1}\sigma^2\sim 1$. Hence the radiation from a kink is emitted in a ``disk-shape" set of directions of solid angle. Then $\mathbf{J}_{\omega}$ becomes
\begin{align}
\label{eq:jmuk_kink}
 \mathbf{J}_{\omega}(\mathbf{k}) 
 \sim \frac{2\mathcal{I}}{L}| t_{*}||\sigma_*| |\Delta\mathbf{x}{'}_{+}| ~,
\end{align}
where $|\Delta\mathbf{x}{'}_{+}|$ denotes the difference between $\mathbf{x}{'}_{+(0+)}$ and $\mathbf{x}{'}_{+(0-)}$. To apply Eq. \eqref{eq:tsigmakink}, we can derive
\begin{equation}
\label{eq:Jomegakink}
 \mathbf{J}_{\omega} \sim \frac{\mathcal{I}}{\omega^{4/3} L^{1/3}} |\Delta\mathbf{x}{'}_{+}| ~,
\end{equation}
and according to Eq. \eqref{eq:powern}, 
\begin{equation}
P_{\omega}=\frac{\mathcal{I}^2}{\omega}\Psi_{+},
\label{eq:kinkp}
\end{equation}
where $\Psi_{+} = |\Delta\mathbf{x}{'}_{+}|$ denotes the sharpness of $\mathbf{x}{'}_{+}$. For a loop with $N$ kinks and sharpness $\Psi_{+}$, the total radiation can be calculated to be
\begin{equation}
 P^{k}_{\gamma} \sim \mathcal{I}^2 N ~ \Psi_{+} ~ \ln \Big( \frac{\omega_{max}}{\omega_{min}} \Big) ~.
\end{equation}
Here $\omega_{max} \sim \sqrt{\mu}$, which is determined by the order of the inverse width of the string. The low frequency cutoff $\omega_{min}$ is set by the validity of the calculation leading to Eq. \eqref{eq:Jomegakink}, which is estimated as $\omega_{min} \sim (L/N)^{-1}$~\cite{Cai:2012zd}.

In the analyses of the radiation event rate in the following sections, we shall take $N \Psi_{+} \sim 1$, which implies that neither the number of kinks nor the sharpness of discontinuity would induce an order-of-magnitude change to the radiations from kinks. With this assumption, we can see that the power emitted from cusps always dominates over that of kinks in total. However, since the kinks can emit radiation in a disk-shape region other than a narrow beam for the case of cusps, it is reasonable to expect that the event rate for kinks would be larger than that of cusps by a factor of $1/\theta \sim (\omega L)^{1/3}$ in certain parameter choices. Also we point out that the calculations above hold in the same way for the case of $\mathbf{x}{'}_{-}$ being discontinuous as well.

\subsection{Kink-kink collision}\label{kk}

In the situation of kink-kink collisions, two opposite-moving kinks meet each other and thus both $\mathbf{x}{'}_{+}$ and $\mathbf{x}{'}_{-}$ are discontinuous. In this case, both $|t_{*}|$ and $|\sigma_{*}|$ are constrained to the order of $\omega^{-1}$, the same as $|t_{*}|$ for kinks. And now $\mathbf{n}$ can be at any direction. The current for the kink-kink collisions becomes
\begin{align}
\label{eq:jmuk_kk}
 \mathbf{J}_{\omega}(\mathbf{k}) & \sim \frac{2\mathcal{I}}{L}| t_{*}||\sigma_*| \mathcal{O}(|\Delta\mathbf{x}{'}_{+}|, |\Delta\mathbf{x}{'}_{-}|) \nonumber\\
 & \sim \frac{2\mathcal{I}}{L\omega^2}\mathcal{O}(|\Delta\mathbf{x}{'}_{+}|, |\Delta\mathbf{x}{'}_{-}|) ~.
\end{align}
Accordingly, the power emitted from the kink-kink collisions at the frequency $\omega$ takes the form
\begin{equation}
\label{eq:kkp}
 P_{\omega}=\frac{\mathcal{I}^2}{\omega^2L}\Psi ~,
\end{equation}
where $\Psi$ denotes $\mathcal{O}(|\Delta\mathbf{x}{'}_{+}|^2,|\Delta\mathbf{x}{'}_{+}||\Delta\mathbf{x}{'}_{-}|,|\Delta\mathbf{x}{'}_{-}|^2)$.

To integrate Eq. \eqref{eq:kkp} over $\omega$ and to assume that there are $N$ left- and right-moving kinks with sharpness $\Psi$, the total power released from the kink-kink collisions is given by
\begin{equation}
 P^{kk}_{\gamma} \sim \frac{\mathcal{I}^2 N \Psi}{\omega_{min} L} \sim \mathcal{I}^2 \Psi ~.
\end{equation}
Under the assumption that $N \Psi \sim 1$, one can easily find that the radiation power from cusps dominate over the other two structures. Eventually, the total electromagnetic radiations from loops of SCSs can be calculated as
\begin{equation}
 P_{\gamma} = P^{c}_{\gamma} + P^{k}_{\gamma} + P^{kk}_{\gamma} ~.
\end{equation}

\section{Event rate of radio signals} \label{sec:eventrate}

As we have derived the spectrum of electromagnetic radiations emitted from various structures of cosmic strings, it is necessary to translate it to the event rate of observational signals associated with experimental instruments. To carry out this process, we need first to investigate the lifetime and number density of these strings that will be presented in the subsection below.

\subsection{String lifetime and number density}

We note that cosmic strings can always emit gravitational radiation as well as electromagnetic radiation. Thus, one can estimate the lifetime of a typical string loop as follows.

The gravitational radiation power takes the form
\begin{equation}
\label{eq:gravi}
 P_g = \Gamma_g G \mu ~,
\end{equation}
where the coefficient takes $\Gamma_g \approx 100$ \cite{Vachaspati:1984gt}. Therefore, the string lifetime can be expressed as follows:
\begin{equation}
 \tau = \frac{\mu L}{P_g+P_{\gamma}} = \frac{L}{\Gamma G \mu} ~,
\end{equation}
where $P_{\gamma} \approx P^{c}_{\gamma}$ and
$\Gamma = {(P_g+P^{c}_{\gamma})}/{(G\mu^2)}.$
As a result, for a string with the initial length $L_i$ at the initial moment $t_i$, its length would evolve as
\begin{equation}
\label{eq:looplength}
 L(t) = L_i - \Gamma G{\mu}(t-t_i) ~,
\end{equation}
with $t \geqslant t_i$.

We are interested in the event rate of radio transients from string loops that exist in the matter dominated phase of the universe. Based on the scaling regime of the evolution model for cosmic strings \cite{Vilenkin:2000jqa} and taking into account the evolution equation for the string length \eqref{eq:looplength}, for loops in the matter area, the number density of string loops can be expressed in terms of the redshift $z$ and the length $L$ as \cite{Cai:2012zd}
\begin{equation}
 dn(L,z) \simeq \frac{C_L(z)(1+z)^6}{t_0^2[(1+z)^{3/2}L + \Gamma G\mu t_0]^2} dL ~,~~ (z<z_{eq}) ~,
\end{equation}
with
\begin{equation}
 C_L(z) = 1 + \frac{t_{eq}^{1/2} (1+z)^{3/4}}{\sqrt{(1+z)^{3/2}L + \Gamma G \mu t_0}} ~.
\end{equation}

\subsection{Burst event rate}

The burst event rate from string loops of length $L$ at redshift $z$ with $N$ left- and right-moving kinks of typical sharpness $\Psi$ per unit volume can be expressed in a general form by \cite{Cai:2012zd}
\begin{equation}
 d \dot {\mathcal{N}}(L,z) \simeq \frac{N^p {\theta_{\nu_0}}^{-{3m}}}{L(1+z)} dn(L,z)dV(z) ~,
\end{equation}
and we have $(p=0,~ m=-2/3)$ for cusps, $(p=1,~ m=-1/3)$ for kinks, and $(p=1,~ m=0)$ for kink-kink collisions, respectively. The angle $\theta_{\nu_0}\sim(\omega L)^{-1/3}$ is the emitted angle of radiations, where $\nu_0$ is the observed frequency. Due to the cosmological expansion, the emitted frequency $\nu_{e}$ is related to $\nu_0$ by the standard redshift relation through
\begin{equation}
 \nu_{e} = \nu_{0}(1+z) ~.
\end{equation}
The physical volume element $dV(z)$ in the matter era is given by
\begin{equation}
 dV(z) = 54\pi t_0^3 [(1+z)^{1/2}-1]^2(1+z)^{-11/2}dz ~,
\end{equation}
and hence, the burst event rate can be written as
\begin{align}
 d \dot {\mathcal{N}}(L,z) \simeq & A ~ N^p (t_0 \nu_0)(\nu_0 L)^{m-1} C_L(z) \nonumber\\
 & \times \frac{(1+z)^{m-1/2}[\sqrt{1+z}-1]^2}{[(1+z)^{3/2}L + \Gamma G \mu t_0]^2} dLdz ~,
\end{align}
with the prefactor $A\sim 50$, which can be determined numerically.

\subsection{Burst flux and duration}\label{subsec:duration}

We have denoted the burst event rate as a function of loop length $L$ and redshift $z$. In the perspective of observations, it is necessary to reformulate it in terms of the energy flux per frequency interval $S$ and duration $\Delta$ of the burst.
To derive the constraint upon the model parameters, in the following we study the transformation from variables $(L, z)$ to $(S, \Delta)$ with $\mathcal{I}$ and $G\mu$ undetermined, and accordingly, the event rate of radio transients, which originally is a function of the loop length $L$ and the redshift $z$, can then be expressed as a function of the observed energy flux $S$ and the observed duration $\Delta$ with a prefactor depending on $\mathcal{I}$ and $G\mu$. In the model of SCSs, the parameter $\mathcal{I}$ depends not only on the loop length $L$ but also on the energy scale of the phase transition that was expected to occur in the very early universe such that these strings could have been generated. In this regard, it can be treated as a free parameter. Therefore, we are allowed to tune the values of $\mathcal{I}$ and $G\mu$ to accommodate with the FRB observations.

We show how $\Delta$ is related with redshift $z$ and the intrinsic frequency $\nu_0$, meanwhile for convenience, we change $d \dot {\mathcal{N}}(L,z)$ from variables $(L,z)$ to $(S,z)$.
The observed duration consists of both the observed intrinsic duration $\Delta t$ and the time broadening induced by the cosmological medium scatterings $\Delta t_s$:
\begin{equation}
\label{eq:Delta}
 \Delta = \Delta t + \Delta t_s ~.
\end{equation}
The intrinsic duration of bursts $\Delta t_{\rm inert}$ measured at the center-of-mass frame of strings is described by Eq.~\eqref{eq:tsigmacusp} for cusps and Eq.~\eqref{eq:tsigmakink} for kinks or kink-kink collisions. This intrinsic pulse is delayed by cosmic expansion and relativistic Doppler effects. Following \cite{Babul:1987lza}, the Lorentz factor at cusps is found to be $\gamma\sim (\omega L)^{1/3}$. In the center-of-mass frame of cosmic strings, due to the relativistic motion of cusps, the pulse width $\Delta t_{\rm inert}$ is then given by $\Delta t_{\rm inert} = \gamma\Delta t_{\rm cusp}$. The observed pulse width is associated with $\Delta t_{\rm cusp}$ through $\Delta t_{\rm obs} =\gamma \Delta t_{\rm cusp}$ if the source is moving away from the observer, $\Delta t_{\rm obs} = \Delta t_{\rm cusp}/\gamma$ if the source is moving toward the observer. Meanwhile, the intrinsic frequency $\omega$ is related to the emitted frequency at the observer's reference frame $\nu_e$ as follows. $\nu_e = \omega/ \gamma$ when the source moves away from the observer, and, $\nu_e = \omega \gamma$ when the source moves toward the observer. In addition, the cosmological expansion can yield the redshift effect upon the frequency via $\nu_o = \nu_e/(1+z)$. By taking into account all the above effects, we can write down the observed intrinsic duration $\Delta t$ to be
\begin{equation}
\Delta t = \frac{L^{1/2}}{\big[ \nu_0(1+z) \big]^{1/2}} ~,
\end{equation}
if the cusp moves away from the observer; and in the opposite case, we have
\begin{equation}
\Delta t = \frac{L^{3/8}}{ \big( \nu_0(1+z) \big)^{3/4}} ~,
\end{equation}
if the cusp moves towards the observer.
We note that the above analysis does not imply that the FRB production from cusps could be related to the motions of strings, but that the observed pulse width of the radiation is dependent of the Lorentz factor due to the fact that the cusps are relativistic. Moreover, the motions of strings would not affect the property that the radiation from cusps is sharply beamed along its forward direction, which can be seen from \eqref{eq:Thetacusp}.

When computing the event rate, we need to take both two cases into consideration. In this regard, the ratio between the backward source and forward source can be treated as a free parameter. However, since the astronomical observations are still far from precise data fitting so far, an estimation of this ratio would be of little significance.
Therefore, we would like to focus our interest only on the backward sources in the detailed estimation. Additionally, in the case of kinks, due to the extremely short intrinsic duration, even after cosmic expansion and relativistic delay, the observed intrinsic duration is still much smaller than the characteristic time of fast radio bursts, and thus, we will omit the observed intrinsic time duration for kinks.

The observed fast radio bursts are all from extra-galactic sources. Accordingly, the scatterings of these bursts are expected to be contributed by both intergalactic medium(IGM) and interstellar medium(ISM). The time broadening effect due to ISM can be related to the dispersion measures of ISM ${\rm DM}_{\rm ISM}$ using an empirical function as follows \cite{Bhat:2004xt}:
\begin{align}
 \Delta t_{\rm ISM} = & -6.5 + 0.15 ~ {\rm log}_{10} ({\rm DM}_{\rm ISM}) \nonumber\\
 & + 1.1 ~ \big( {\rm log}_{10} ({\rm DM}_{\rm ISM}) \big)^2 - 3.9 ~ {\rm log}_{10}(\nu_0) ~,
\end{align}
where ${\rm DM}_{\rm ISM}$ is often treated as a constant, which roughly equals $95~pc/cm^3$.
The time broadening function due to the IGM scattering is rescaled as three orders of magnitude smaller than that due to ISM \citep{Lorimer:2013roa, Caleb:2015uuk}
\begin{align}
 \Delta t_{\rm IGM} = & -9.5 + 0.15 ~ {\rm log}_{10} ({\rm DM}_{\rm IGM}) \nonumber\\ 
 & + 1.1 ~ \big( {\rm log}_{10} ({\rm DM}_{\rm IGM}) \big)^2 - 3.9 ~ {\rm log}_{10}(\nu_0) ~,
\end{align}
where ${\rm DM}_{\rm IGM}$ is a function of $z$ as follows \cite{Ioka:2003fr}:
\begin{align}
 {\rm DM}_{\rm IGM}(z) = \frac{3cH_0\Omega_b}{8\pi G m_p}\int^z_0 \frac{(1+z')dz'}{\sqrt{\Omega_m(1+z)^3 +\Omega_{\Lambda}}} ~.
\end{align}
As a result, the time duration broadening due to scattering can be determined by
\begin{equation}
 \Delta t_s(z) = \Delta t_{\rm ISM} + \Delta t_{\rm IGM}(z) ~.
\end{equation}

The energy flux per frequency interval $S$ of a string loop observed at a distance $r(z)$ is obtained through dividing the power radiated per unit frequency per unit solid angle by $r(z)^2$. The power per unit frequency emitted by cusps, kinks and kink-kink collisions can be found in Eqs. \eqref{eq:cuspp}, \eqref{eq:kinkp}, \eqref{eq:kkp}, respectively. The solid angles of cusps and kinks are provided in Eqs. \eqref{eq:Thetacusp}, \eqref{eq:Thetakink}, respectively. We mention that the expressions of the power are averaged over a frequency interval $T=L/2$, and thus, we need to multiply them by $T$ and then divide them by the burst duration $\Delta$ to get the energy flux. After that, the observed energy flux $S$ for cusps, kinks and kink-kink collisions can then be derived as
\begin{equation}
\label{eq:flux}
 S \approx \frac{L^2 \mathcal{I}^2}{r(z)^2 \Delta} ~ \frac{\Psi^p}{\big[ \nu_0 L(1+z) \big]^{q}} ~,
\end{equation}
where $(p=0,~ q=0)$ for cusps; $(p=1,~ q=2/3)$ for kinks; and $(p=1,~ q=2)$ for kink-kink collisions, respectively. The proper distance $r(z)$ in the matter dominated flat universe is given by
\begin{equation}
\label{eq:rz}
 r(z) = 3 t_0 \frac{\big[ (1+z)^{1/2}-1 \big]}{(1+z)^{1/2}} ~.
\end{equation}

Note that, with the help of Eq. \eqref{eq:flux}, we are able to express $S$ as a function of $L$, and thus, can further translate the burst event rate from a function of $(L,z)$ to that of $(S,z)$. The previous estimation of $\Delta$ explicitly shows that $\Delta$ mainly depends on $z$. Therefore, $dLdz = |\partial L/\partial S|dSdz$. In addition, we also comment that this transformation can only work for the cases of cusps and kinks. In the situation of kink-kink collisions, however, as $S$ is also $L$-independent we need to treat this case separately as will be shown later.

For cusps and kinks, we can determine from Eq. \eqref{eq:flux} that
\begin{equation}
 \bigg{|} \frac{\partial L}{\partial S} \bigg{|} = \frac{L}{(2-q)S} ~,
\end{equation}
where $L$ can be written as
\begin{equation}
 L = \Big{[} \frac{\nu_0^{q-1}}{\Psi^p} \frac{S}{\mathcal{I}^2} r^2(z)(1+z)^q x \Big{]}^{1/(2-q)} ~,
\end{equation}
with $x \equiv \Delta \nu_0.$

Eventually we can find
\begin{equation}
 dLdz = \frac{L}{(2-q)S}dSdz ~.
\end{equation}
With this relation, the burst event rate now can be rewritten as
\begin{equation}
\label{eq:eventratecuspkink}
 d \dot {\mathcal{N}}(S,z) \simeq \frac{At_0 \nu_0^m N^p}{(2-q) S} [L(x,S)]^m f_m(x,S) dSdz ~,
\end{equation}
where
\begin{equation}
 f_m(x,S) = C_L(z)\frac{(1+z)^{m-1/2}[\sqrt{1+z}-1]^2 }{\big[ (1+z)^{3/2}L(x,S) + \Gamma G\mu t_0 \big]^2} ~.
\end{equation}

Now we have the general expression of the burst event rate for cusps and kinks. As we mentioned previously, however, for kink-kink collisions both $S$ and $\Delta$ depend on $z$ only. It implies that $S$ and $\Delta$ are not independent from each other any more. As a result, for kink-kink collisions, we shall estimate the burst event rate as a function of $z$ only. Correspondingly, $S$ and $\Delta$ can be derived from Eqs. \eqref{eq:flux} and \eqref{eq:Delta}. Note that from $S$ and $\Delta$ we cannot determine $L$, and hence we need to integrate the burst event rate $d \dot {\mathcal{N}}(L,z)$ over $L$ to have the total amount $d \dot {\mathcal{N}}(z)$ as follows:
\begin{equation}
\label{eq:eventratekk}
 d \dot {\mathcal{N}}(z) = \frac{A N t_0 t_{eq}^{1/2} (1+z)^{1/4}[\sqrt{1+z}-1]^2}{(\Gamma G \mu t_0)^{5/2}}  dz ~.
\end{equation}

\section{Numerical computation and the FRB data fitting} \label{sec:numer&data}

So far we have derived the expressions of the burst event rate coming from SCSs. Accordingly, we now are able to confront the theoretical predictions of SCSs with the observational data of FRB such that model parameters of SCSs can be well constrained. Note that, from the perspective of a theoretical description as performed in previous sections, we have five parameters that characterize the application of SCSs in explaining the FRB data, which are: $\mathcal{I}$, $G\mu$, $\nu_0$, $\Delta$, $S$, respectively. The first two of these parameters are associated with the background theory and the latter three are connecting to observations.

Before doing the detailed data fitting of the FRB, we would like to recall that the study on the CMB background anisotropies and the stochastic gravitational wave background has imposed an upper limit on the string tension, i.e., $G\mu < 1.3 \times 10^{-7}$ \cite{Ade:2013xla}. Moreover, for SCSs the parameter space of $\mathcal{I} > 10^4 ~ {\rm GeV}$ and $10^{-19} < G\mu < 10^{-7}$ has been ruled out due to the constraint on spectral distortions of the CMB photons \cite{Kogut:2011xw}. Another limit on $\mathcal{I}$ is related with $G\mu$ under the requirement that the dominant process of energy loss is due to the electromagnetic radiations rather than the gravitational radiations; otherwise, the gravitational radiations might have been detected accompanied with the electromagnetic signals observed in astronomical experiments. Accordingly, given $P^c_{\gamma} = P_g$, the critical current $\mathcal{I}_*$ can be determined as follows \cite{Cai:2012zd}:
\begin{equation}
 \mathcal{I}_* = \frac{\Gamma_g}{\Gamma^c_{\gamma}} G \mu^{3/2}\ \simeq 10^{20} \times {G\mu}^{3/2}\ {\rm GeV} ~.
\end{equation}

Consequently, for a given $G\mu$, one expects that the value of $\mathcal{I}$ is larger than a critical value $\mathcal{I}_*$ derived above to ensure that the major channel for a superconducting string to release energy is via electromagnetic radiations. Using Eqs. \eqref{eq:eventratecuspkink} and \eqref{eq:eventratekk}, one can see that for each pair of $(G\mu, \mathcal{I})$ the event rate of radio bursts can be expressed as a function of redshift. In this way we are able to fit the event rate function predicted by the theory of SCSs with the normalized observational data. With this purpose, we also need to investigate the threshold flux and the threshold pulse width from the perspective of observations.

Based on the analyses in Refs. \cite{Caleb:2015uuk, Bera:2016qfe}, one finds that the signal-to-noise ratio of each event can be determined by
\begin{equation}
 S_*/N = \frac{S_*G_{sys}\sqrt{\Delta B N_{pol}}}{T_{sys}} ~,
\end{equation}
where $G_{sys}$ is the system gain and approximately equals $0.7\ K/Jy$, $B$ is the bandwidth and approximately equals $340 ~ {\rm MHz}$, $N_{pol}$ is the polarization numbers that takes the value $2$, and, $T_{sys}$ is the temperature of the system which is set as $21 \ K$. All these values of the system parameters are taken from the setup of the Parkes multi-beam receiver \citep{Caleb:2015uuk}. Note that $S_*$ is the threshold flux for the detection and we choose the value of $S_*/N$ to be $10$. The pulse width $\Delta$ can be determined by Eq.~\eqref{eq:Delta} derived in the previous section. Additionally, the threshold pulse width required for the detection is set to be $0.1\ ms$.

After having settled down the system restrictions, we are able to calculate $d\dot{\mathcal{N}}(z)/dz$ case by case for cusps, kinks and kink-kink collisions. The frequency band of the Parkes multi-beam receiver is set as $\nu_0\sim[1.182, 1.522]~{\rm GHz}$. We integrate Eq. \eqref{eq:eventratecuspkink} for $\nu_0$ in the Parkes bandwidth, and there is $S\sim[10^{-1},10^{2}]~{\rm Jy}$, which is consistent with the flux range of the detected events.

The event rate of radio bursts with $S<10^{-1}{\rm Jy}$ has little contribution to the total amount, since that their flux is lower than the threshold flux $S_*$. In addition, the event rate with $S>10^{2}~{\rm Jy}$ is also much suppressed and has almost no contribution to the total event rate. This is because the upper bound of $S$ is quite well restrained by the length of string loops and the threshold observational pulse width. Eq.~\eqref{eq:flux} shows that, for both cusps and kinks, $S$ has a power-law correlation with $L$ and an inverse relation with $\Delta$. From Eq.~\eqref{eq:looplength}, we can find an upper bound of the loop length: $L_i - \Gamma G \mu (t-t_{eq})$, where $L_i$ is the string length at some initial moment of the cosmic evolution, and $t$ is the cosmic time related to the redshift where the observed string would be located. Since $L_i$ is much larger than $\Gamma G\mu t_0$, this restriction upon the string length is often negligible. For cusps, there is another limit from Eq.~\eqref{eq:cusplimitL} that requires $L<{\mu^{3/2}}/{\mathcal{I}^3\omega_{max}}$, and thus a radio signal from the cusp has to satisfy this inequality. For kink-kink collisions, we use Eq.~\eqref{eq:eventratekk} to calculate the event rate straightforwardly as there is no further constraint of the loop length in this case. However, we will show later that the flux from kink-kink collisions is typically much smaller than the threshold flux $S_*$ required by the present detection sensitivity. Therefore, we conclude that there is very little chance to observe such an occurrence by the present detection ability in radio astronomy.

\begin{widetext}
\begin{center}
\begin{table}
\begin{tabular}{cccccc}
\hline
Event & Redshift & $\rm{DM[cm^{-3} pc]}$ & $\rm{W_{obs}[ms]}$ & $\rm{S_{peak, obs}[Jy]}$ & $\rm{F_{obs}[Jy\ ms]}$\\
\hline
FRB010125& 0.57& 790& $\rm{9.40}_{-0.20}^{+0.20}$& 0.30& 2.82\\
FRB010621& 0.19& 745& 7.00& 0.41& 2.87\\
FRB010724& 0.28& 375& 5.00& $>30.00_{-10.00}^{+10.00}$& $>$150.00\\
FRB090625& 0.72& 899.55& $\rm{1.92}_{-0.77}^{+0.83}$& $1.14_{-0.21}^{+0.42}$& $2.19_{-1.12}^{+2.10}$\\
FRB110220& 0.76& 944.38& $\rm{5.60}_{-0.10}^{+0.10}$& $1.30_{-0.00}^{+0.00}$& $7.28_{-0.13}^{+0.13}$\\
FRB110626& 0.56& 723.0& 1.40& 0.40& 0.56\\
FRB110703& 0.89& 1103.6& 4.30& 0.50& 2.15\\
FRB120127& 0.43& 553.3& 1.10& 0.50& 0.55\\
FRB121002& 1.3& 1629.18& $\rm{5.44}_{-1.20}^{+3.50}$& $\rm{0.43}_{-0.06}^{+0.33}$& $2.34_{-0.77}^{+4.46}$\\
FRB130626& 0.74& 952.4& $\rm{1.98}_{-0.44}^{+1.20}$& $\rm{0.74}_{-0.11}^{+0.49}$& $1.47_{-0.50}^{+2.45}$\\
FRB130628& 0.35& 469.88& $\rm{0.64}_{-0.13}^{+0.13}$& $\rm{1.91}_{-0.23}^{+0.29}$& $1.22_{-0.37}^{+0.47}$\\
FRB130729& 0.69& 861& $\rm{15.61}_{-6.27}^{+9.98}$& $\rm{0.22}_{-0.05}^{+0.17}$& $3.43_{-1.81}^{+6.55}$\\
FRB131104& 0.59& 779& 2.08& 1.12& 2.33\\
FRB140514& 0.44& 562.7& $2.80_{-0.70}^{+0.35}$& $\rm{0.47}_{-0.08}^{+0.11}$& $1.32_{-0.50}^{+2.34}$\\
FRB150418& 0.49& 776.2& $0.80_{-0.30}^{+0.30}$& $\rm{2.20}_{-0.30}^{+0.60}$& $1.76_{-0.81}^{+1.32}$\\
\hline
\end{tabular}
\caption{FRB catalog as detected by Parkes \cite{Petroff:2016tcr}.}
\label{table:frbcatalog}
\end{table}
\end{center}
\end{widetext}

Recall that the FRB parameters in association with the cosmic string model are: the derived redshift $z$, the dispersion measures DM (unit: $\rm{cm^{-3} pc}$), the pulse width $\rm{W_{obs}}$ (unit: ms), the energy flux $\rm{S_{obs}}$ or $\rm{F_{obs}}$~(unit: Jy or Jy ms), respectively. In Table.~\ref{table:frbcatalog} we list 15 FRB events detected by the Parkes multi-beam receiver \cite{Petroff:2016tcr}. Among them, the redshift of the furthest source is about $z=1.3$. For convenience of numerical analysis, we separate the regime of $z\sim[0, 1.4]$ into 7 bins and count the event number $\Delta N_{obs}/\Delta z_{bin}$ in each bin. The error bars of observational data are set as ${\rm e}_{\rm obs} = \sqrt{\Delta N_{obs}/\Delta z_{bin}}$. We then normalize the data so that the sum of 7 bins satisfies
\begin{equation}
 \sum {\rm n}_{\rm obs} \Delta z_{bin} = \sum(\Delta N_{obs}/\Delta z_{bin}) \Delta z_{bin} = 1 ~,
\end{equation}
where ${\rm n}_{\rm obs}$ stands for the normalized event rate per redshift bin. To compare theoretical prediction with it, for each $(\mu, \mathcal{I})$, we normalize the $d\dot {\mathcal{N}}(z)$ function so that the normalized theoretical event rate per redshift ${\rm n}_{\rm th}$ satisfies $\int{\rm n}_{\rm th} dz = \int (d\dot {\mathcal{N}}(z)/dz) dz = 1$. We use the $\chi^2$ value to denote the quality of fitting between these two normalized data settings, where $\chi^2$ is defined as
\begin{equation}
\label{eq:fitting}
 \chi^2 = {\sum_{i=1}^{7}} \frac{({\rm n}_{\rm obs} - {\rm n}_{\rm th})^2}{{{\rm e}_{\rm obs}}^2} ~.
\end{equation}

\begin{figure}
\includegraphics[width=0.4\textwidth]{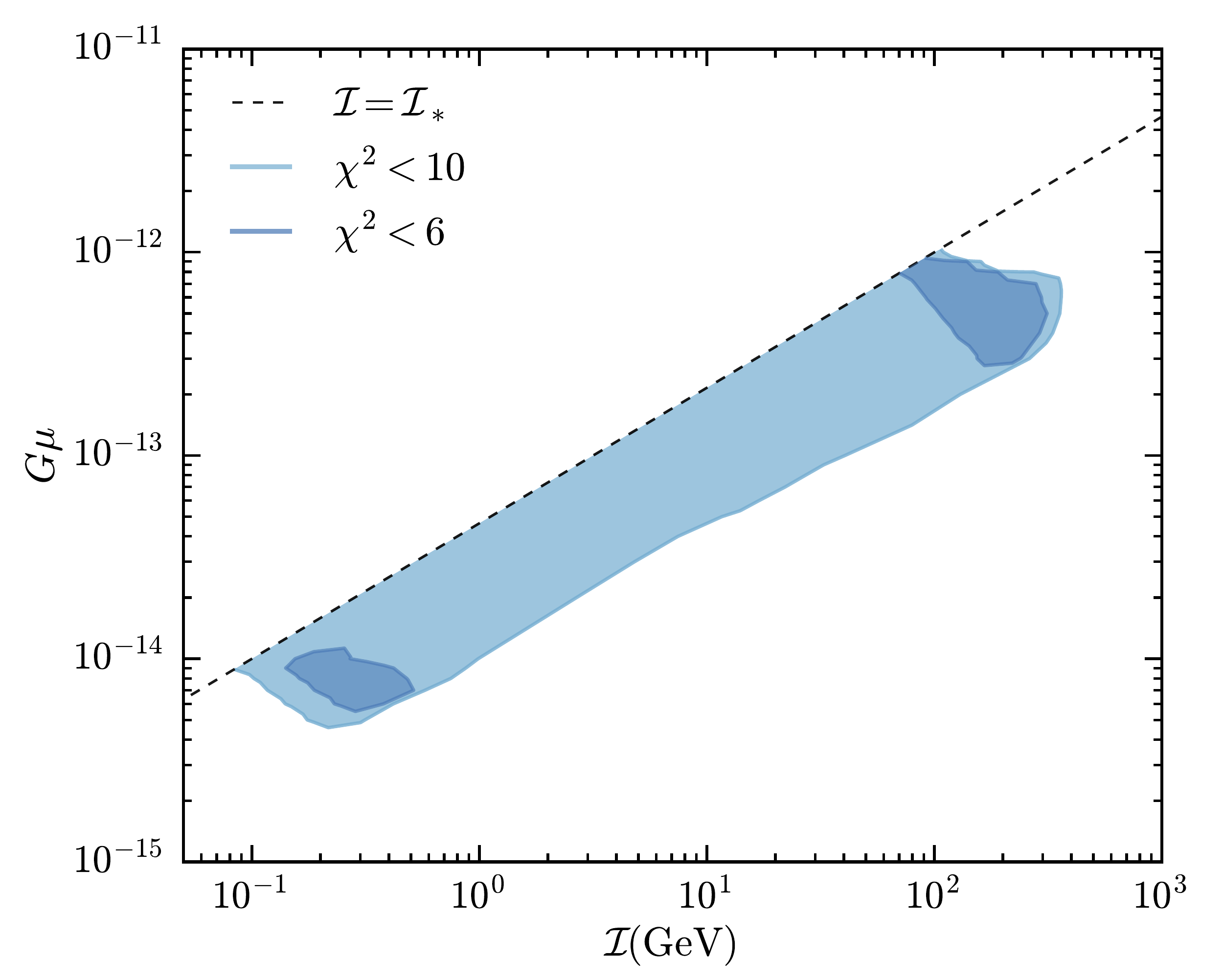}
\includegraphics[width=0.4\textwidth]{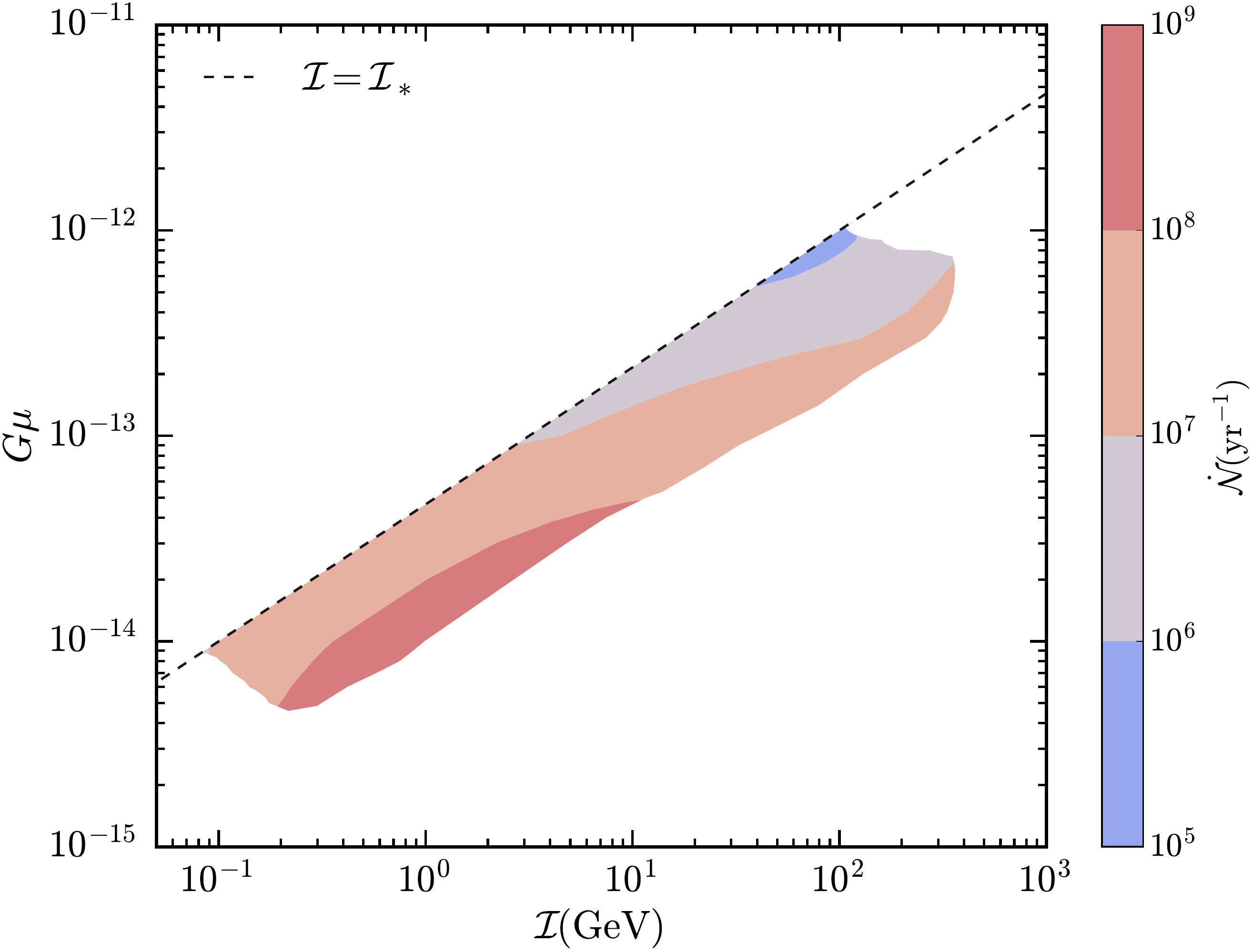}
\caption{
Upper: The $\chi^2$ contour for the parameter space of $G\mu$ and $\mathcal{I}$.
Lower: The distribution of the burst event rate $\dot{\mathcal{N}}$ within the parameter space with $\chi^2 < 10$.
The black dashed line in the upper panel represents for the boundary where $\mathcal{I}=\mathcal{I}_*$.
The regime below this line corresponds to where electromagnetic radiations dominate over gravitational ones.
}
\label{fig:chi}
\end{figure}

Afterwards, we compute the value of $\chi ^2$ for $(\mu, \mathcal{I})$ pairs that fit the above settings, with the viable result being presented in the upper panel of Fig.~\ref{fig:chi}.
The black dashed line in Fig.~\ref{fig:chi} denotes the boundary where $\mathcal{I}$ equals the critical value at which the gravitational radiation is comparable with electromagnetic radiation.
We are interested in the parameter space dominated by electromagnetic radiation that is below the black dashed line.
In this area, the light blue contour corresponds to the regime with $\chi^2 < 10$. If we further decrease the value of $\chi^2$ to be $\chi^2<6$, we interestingly find that there exist two separate regimes allowed by the FRB data, which are shown as the dark blue contours in the upper panel of Fig.~\ref{fig:chi}.
Here, we would like to emphasize that, due to the very limit sample of the FRB data, the statistics of the data fitting remains very poor and thus the confidence level is quite low.
According to our numerical analysis, the significance is approximately $1.3 \sigma$ for the contour with $\chi^2 < 10$.
From the numerical estimation, we observe that the allowed range is roughly within $G\mu\sim[10^{-14}, 10^{-12}]$, and, $\mathcal{I}\sim[10^{-1}, 10^3]~{\rm GeV}$.
Given the observationally allowed parameter space as shown in the upper panel of Fig.~\ref{fig:chi}, we numerically derive the corresponding distribution of the burst event rate per year in the lower panel of the same figure.
Accordingly, we find that the event rate $\dot{\mathcal{N}}$ within the colored region is between $10^5 \sim 10^9~{\rm yr}^{-1}$. This result implies that the observation of the radio bursts from SCSs is promising in various experiments of radio astronomy.

\begin{figure}
\includegraphics[width=0.45\textwidth]{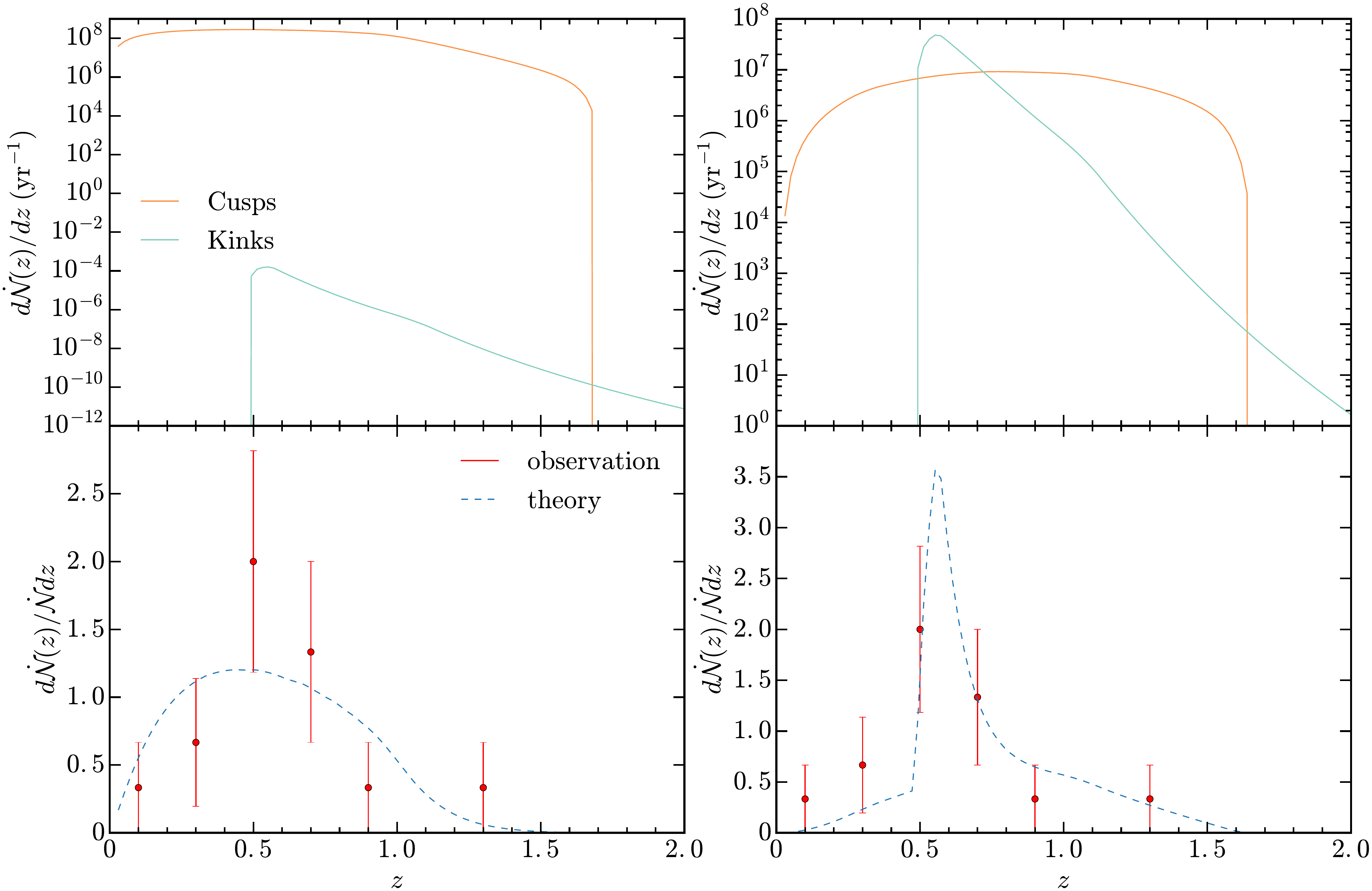}
\caption{
Numerical estimations of the event rate of radio bursts emitted from SCSs with two different choices of model parameters as well as their comparison with the observational data of FRB. In the left panel, the values of parameters are provided in Eq.~\eqref{eq:solutionA}, while, for the right panel the values are given by Eq.~\eqref{eq:solutionB}. In the upper two plots the event rates were counted for one year and in the lower plots we presented the normalized event rates.
}
\label{fig:bestfitting}
\end{figure}

Note that best fitting contours could lie on two sides of the parameter space as observed in the upper panel of Fig~\ref{fig:chi}. For each contour, one can derive a best fitting spot in the parameter space where $\chi^2$ is the lowest. We numerically investigate the event rate as a function of the redshift for both two ``best fitting'' spots as shown in Fig.~\ref{fig:bestfitting}. In this figure, the model parameters are determined as follows. For the left panel, we have 
\begin{equation}
\label{eq:solutionA}
 G\mu = 5\times 10^{-15}, \mathcal{I} = 0.313~{\rm GeV}
\end{equation}
with $\chi^2=4.72$ and the corresponding event rate is found to be $\dot{\mathcal{N}}\sim 2.3\times 10^8$~${\rm yr}^{-1}$; moreover, for the right panel, the model parameters take the values 
\begin{equation}
\label{eq:solutionB}
 G\mu = 3\times 10^{-13}, \mathcal{I} = 236.28~{\rm GeV}
\end{equation}
with $\chi^2=3.88$ and the event rate takes $\dot{\mathcal{N}}\sim 1.6\times 10^7$~${\rm yr}^{-1}$.

From the numerical estimations presented in Fig.~\ref{fig:bestfitting}, we can see that the event rate of the above two cases are slightly higher than the estimation in other work, namely, $\dot{\mathcal{N}} \sim 10^{6}~{\rm yr}^{-1}$ as was estimated in Ref.~\cite{Caleb:2015uuk}. For each case of our investigation, in the upper panel, the solid orange curves show the event rate emitted from the cusps and the solid green curves depict the event rate arisen from the kinks. There is no signal from the kink-kink collisions shown in the figure for the reason that either their flux is much lower than the threshold of the experimental detection or their event rate is far below detectability. In the lower panel, the blue dashed line denotes the normalized total event rate, and, the red points are normalized observational data. So far, the observed events exhibit a peak value around the redshift $z=0.5$. In the case shown in the left panel, radio bursts due to string cusps make dominant contribution throughout the whole redshift range $z\sim(0,1.6)$; while, in the right case, radio burst signals from kinks can become dominant around the redshift $z=0.5\sim0.7$, with the corresponding amplitude being about ten times larger than that from cusps. This interesting local effect happens to overlap with the bulging profile of the observational data as shown in the lower right panel. Therefore, if the local feature of high event rate of radio bursts within the range of $z=0.5\sim0.7$ were confirmed in the accumulated observations with high precision, the theory of SCSs could provide an interesting explanation of this possible phenomenon.

For both two cases, the burst event rate from cusps exhibits an abrupt cutoff at the high redshift regime due to the limitation of the string length. In our detailed calculation, we have constrained the frequency range to be within $1.182\sim1.522$~${\rm GHz}$ in order to be consistent with the experimental parameter of Parkes. As the frequency increases, we see that the cutoff redshift would also increase. For instance, $z\sim 2$ would correspond to $\nu_0\sim 3~{\rm GHz}$. In the low redshift regime, the burst event rate decreases mainly due to the fact that the pulse width is too small. Moreover, we find that the observed pulse width shows very little dependence on the redshift. However, according to the scattering model analyzed in Sec. \ref{subsec:duration}, one expects that the pulse width broadening due to the ISM and IGM effects should be correlated with the redshift. We argue that this issue may be resolved either by a fine tuning of the background theory of SCSs or by a better version of the scattering model.

\section{Conclusion and discussion} \label{sec:discussion}

To summary, we in this paper have applied the observational data of FRB to constrain model parameters appeared in the theory of the superconducting cosmic strings. In particular, based on the method of the order-of-magnitude estimation, we have recomputed the electromagnetic emission power of three kinds of string structures, which are cusps, kinks and kink-kink collisions, respectively. After that, we have demonstrated in Fig.~\ref{fig:chi} that the SCSs models within the parameter space with $G\mu\sim[10^{-14}, 10^{-12}]$ and $\mathcal{I}\sim[10^{-1}, 10^3]~{\rm GeV}$ are consistent with the present FRB observations. By performing detailed numerical calculations, we surprisingly found two possible best fitting spots as estimated by the $\chi^2$ value. One best fitting spot corresponds to the string model with parameters being $G\mu = 5 \times 10^{-15}$ and $\mathcal{I} = 0.313 ~ \rm{GeV}$ and in this case the event rates of the radio bursts are mostly contributed by cusps; the other one corresponds to the model with $G\mu = 3 \times 10^{-13}$ and $\mathcal{I} = 236.28 ~ \rm{GeV}$, in which the contribution of kinks could dominate over that of cusps in the intermediate regime along the redshift. With accumulated high precision observations of radio telescopes in near future, it will be very promising to examine or to rule out the theoretical possibility of using SCSs as sources of cosmological FRB.

Recall that we have pointed out in Sec.~\ref{subsec:duration} that radio bursts can be generated by cusps moving either backwards or towards the observers. Radio bursts from those forward sources can have larger value of the event rate than the backward sources over certain redshift by an order of magnitude. Such differences become small as $G\mu$ increases. It is then natural to consider the phenomenon that the ratio of the backward sources over forward ones would also affect the analyses performed in the present study. As future observations will provide us a more reliable relationship between the event rate and  the redshift, it will also be important to take into account this ratio parameter in the follow-up study.

In addition, among all the detected FRB events, it has been reported that the event FRB 150418 and FRB 150807 are linearly polarized \cite{Keane:2016yyk, Ravi:2016kfj}, and FRB 140514 and FRB 110523 have shown the mixture of linear and circular polarizations \cite{Petroff:2014taa, Masui:2015kmb}. We note that the circular polarization could be caused by the foreground contamination, such as, the Faraday rotation effect when the radio signals pass through the magnetized region and/or the high-density medium in Milky Way as well as intergalactic environments, or even possibly the host galaxy. According to the study of \cite{Petroff:2014taa}, the event FRB 140514 has shown that the linear polarization is intrinsic to the signal. In the literature there have been attempts in understanding these particular patterns of polarizations \cite{Pen:2015ema}. Importantly, the study of SCSs in \citep{Cai:2011bi} has already demonstrated that radiations from string cusps should be linearly polarized while no specific prediction on the polarization was made for kinks. In this regard, the precise measurement of the circular polarization can help to identify the host environment of FRB sources and the theoretical origin of SCSs. Although these environmental issues remain unclear, so far the present FRB observations yet cannot rule out the models of SCSs. Thus, it deserves to study in detail the potential co-evolution of SCSs and galaxy in the future in order to examine possible connections among the polarization of radio signals, the models of SCSs and the unknown environmental effects.


We end by noting that our present results are also consistent with the present constraints on the parameters of the SCSs models, for instance, as obtained in \cite{Miyamoto:2012ck, Brandenberger:2017uwo}. In fact, our analysis yields more stringent constraints with the starting point that SCSs could play the role of the theoretical origin of cosmological FRB. With this theoretical possibility, the current parameter of the strings could be as low as the GeV scale, and thus, might become of observational interest in high energy phenomenology as well as experiments, such as, the Large Hadron Collider, the International Linear Collider, the Circular Electron Positron Collider and so on.

\section*{Acknowledgments}
We are grateful to Hong Guo, Eray Sabancilar, Daniele A. Steer, Tanmay Vachaspati and Tinggui Wang for valuable communications.
We especially thank Zhi-Qiang Shen for insightful discussion on the frontier of radio astronomy.
J.Y. and K.W. are supported in part by the NSFC Fund for Fostering Talents in Basic Science (No. J1310021).
Y.F.C. is supported in part by the Chinese National Youth Thousand Talents Program, by the CAST Young Elite Scientists Sponsorship Program (2016QNRC001), by the NSFC (Nos. 11421303, 11653002), and by the Fundamental Research Funds for the Central Universities.
Part of numerical simulations are operated on the computer cluster LINDA in the particle cosmology group at USTC.


\begin{thebibliography}{99}

\bibitem{Kibble:1976sj}
  T.~W.~B.~Kibble,
  J.\ Phys.\ A {\bf 9}, 1387 (1976).

\bibitem{Vilenkin:2000jqa}
  A.~Vilenkin and E.~P.~S.~Shellard,
  Cambridge University Press (1994).

\bibitem{Polchinski:2004ia}
  J.~Polchinski,
  hep-th/0412244.

\bibitem{Copeland:2009ga}
  E.~J.~Copeland and T.~W.~B.~Kibble,
  Proc.\ Roy.\ Soc.\ Lond.\ A {\bf 466}, 623 (2010)
  [arXiv:0911.1345 [hep-th]].

\bibitem{Ringeval:2010ca}
  C.~Ringeval,
  Adv.\ Astron.\  {\bf 2010}, 380507 (2010)
  [arXiv:1005.4842 [astro-ph.CO]].

\bibitem{Copeland:2011dx}
  E.~J.~Copeland, L.~Pogosian and T.~Vachaspati,
  Class.\ Quant.\ Grav.\  {\bf 28}, 204009 (2011)
  [arXiv:1105.0207 [hep-th]].

\bibitem{Vachaspati:2015cma}
  T.~Vachaspati, L.~Pogosian and D.~Steer,
  Scholarpedia {\bf 10}, no. 2, 31682 (2015)
  [arXiv:1506.04039 [astro-ph.CO]].

\bibitem{Witten:1984eb}
  E.~Witten,
  Nucl.\ Phys.\ B {\bf 249}, 557 (1985).

\bibitem{Vilenkin:1986zz}
  A.~Vilenkin and T.~Vachaspati,
  Phys.\ Rev.\ Lett.\  {\bf 58}, 1041 (1987).

\bibitem{Garfinkle:1987yw}
  D.~Garfinkle and T.~Vachaspati,
  Phys.\ Rev.\ D {\bf 36}, 2229 (1987).

\bibitem{Sanchez:1988ek}
  N.~G.~Sanchez and M.~Signore,
  Phys.\ Lett.\ B {\bf 219}, 413 (1989).

\bibitem{Sanchez:1990kj}
  N.~G.~Sanchez and M.~Signore,
  Phys.\ Lett.\ B {\bf 241}, 332 (1990).

\bibitem{Tashiro:2012nb}
  H.~Tashiro, E.~Sabancilar and T.~Vachaspati,
  Phys.\ Rev.\ D {\bf 85}, 103522 (2012)
  [arXiv:1202.2474 [astro-ph.CO]].

\bibitem{Jazayeri:2017szw}
  S.~Jazayeri, A.~V.~Sadr and H.~Firouzjahi,
  Phys.\ Rev.\ D {\bf 96}, no. 2, 023512 (2017)
  [arXiv:1703.05714 [astro-ph.CO]].

\bibitem{Tashiro:2012nv}
  H.~Tashiro, E.~Sabancilar and T.~Vachaspati,
  Phys.\ Rev.\ D {\bf 85}, 123535 (2012)
  [arXiv:1204.3643 [astro-ph.CO]].

\bibitem{Hill:1986mn}
  C.~T.~Hill, D.~N.~Schramm and T.~P.~Walker,
  Phys.\ Rev.\ D {\bf 36}, 1007 (1987).

\bibitem{Bhattacharjee:1989vu}
  P.~Bhattacharjee,
  Phys.\ Rev.\ D {\bf 40}, 3968 (1989).

\bibitem{Bhattacharjee:1990js}
  P.~Bhattacharjee and N.~C.~Rana,
  Phys.\ Lett.\ B {\bf 246}, 365 (1990).

\bibitem{MacGibbon:1989kk}
  J.~H.~MacGibbon and R.~H.~Brandenberger,
  Nucl.\ Phys.\ B {\bf 331}, 153 (1990).

\bibitem{Wichoski:1998kh}
  U.~F.~Wichoski, J.~H.~MacGibbon and R.~H.~Brandenberger,
  Phys.\ Rev.\ D {\bf 65}, 063005 (2002)
  [hep-ph/9805419].

\bibitem{Brandenberger:2009ia}
  R.~Brandenberger, Y.~F.~Cai, W.~Xue and X.~m.~Zhang,
  arXiv:0901.3474 [hep-ph].

\bibitem{Berezinsky:2009xf}
  V.~Berezinsky, K.~D.~Olum, E.~Sabancilar and A.~Vilenkin,
  Phys.\ Rev.\ D {\bf 80}, 023014 (2009)
  [arXiv:0901.0527 [astro-ph.HE]].

\bibitem{Lunardini:2012ct}
  C.~Lunardini and E.~Sabancilar,
  Phys.\ Rev.\ D {\bf 86}, 085008 (2012)
  [arXiv:1206.2924 [astro-ph.CO]].

\bibitem{Babul:1987lza}
  A.~Babul, B.~Paczynski and D.~Spergel,
  Astrophys.\ J.\  {\bf 316}, L49 (1987).

\bibitem{Paczynski:1988}
  B. Paczynski,
  Astrophys.\ J.\ {\bf 335}, 525 (1988).

\bibitem{Brandenberger:1993hw}
  R.~H.~Brandenberger, A.~T.~Sornborger and M.~Trodden,
  Phys.\ Rev.\ D {\bf 48}, 940 (1993)
  [hep-ph/9302254].

\bibitem{Plaga:1993hp}
  R.~Plaga,
  Astrophys.\ J.\  {\bf 424}, L9 (1994).

\bibitem{Berezinsky:2001cp}
  V.~Berezinsky, B.~Hnatyk and A.~Vilenkin,
  Phys.\ Rev.\ D {\bf 64}, 043004 (2001)
  [astro-ph/0102366].

\bibitem{Cheng:2010ae}
  K.~S.~Cheng, Y.~W.~Yu and T.~Harko,
  Phys.\ Rev.\ Lett.\  {\bf 104}, 241102 (2010)
  [arXiv:1005.3427 [astro-ph.HE]].

\bibitem{Vachaspati:1984gt}
  T.~Vachaspati and A.~Vilenkin,
  Phys.\ Rev.\ D {\bf 31}, 3052 (1985).

\bibitem{Damour:2000wa}
  T.~Damour and A.~Vilenkin,
  Phys.\ Rev.\ Lett.\  {\bf 85}, 3761 (2000)
  [gr-qc/0004075].

\bibitem{Damour:2001bk}
  T.~Damour and A.~Vilenkin,
  Phys.\ Rev.\ D {\bf 64}, 064008 (2001)
  [gr-qc/0104026].

\bibitem{Damour:2004kw}
  T.~Damour and A.~Vilenkin,
  Phys.\ Rev.\ D {\bf 71}, 063510 (2005)
  [hep-th/0410222].

\bibitem{Vachaspati:2008su}
  T.~Vachaspati,
  Phys.\ Rev.\ Lett.\  {\bf 101}, 141301 (2008)
  [arXiv:0802.0711 [astro-ph]].

\bibitem{Cai:2011bi}
  Y.~F.~Cai, E.~Sabancilar and T.~Vachaspati,
  Phys.\ Rev.\ D {\bf 85}, 023530 (2012)
  [arXiv:1110.1631 [astro-ph.CO]].

\bibitem{Cai:2012zd}
  Y.~F.~Cai, E.~Sabancilar, D.~A.~Steer and T.~Vachaspati,
  Phys.\ Rev.\ D {\bf 86}, 043521 (2012)
  [arXiv:1205.3170 [astro-ph.CO]].

\bibitem{Yu:2014gea}
  Y.~W.~Yu, K.~S.~Cheng, G.~Shiu and H.~Tye,
  JCAP {\bf 1411}, no. 11, 040 (2014)
  [arXiv:1409.5516 [astro-ph.HE]].

\bibitem{Dvorkin:2011aj}
  C.~Dvorkin, M.~Wyman and W.~Hu,
  Phys.\ Rev.\ D {\bf 84}, 123519 (2011)
  [arXiv:1109.4947 [astro-ph.CO]].

\bibitem{Ade:2013xla}
  P.~A.~R.~Ade {\it et al.} [Planck Collaboration],
  Astron.\ Astrophys.\  {\bf 571}, A25 (2014)
  [arXiv:1303.5085 [astro-ph.CO]].

\bibitem{vanHaasteren:2011ni}
  R.~van Haasteren {\it et al.},
  Mon.\ Not.\ Roy.\ Astron.\ Soc.\  {\bf 414}, no. 4, 3117 (2011)
  Erratum: [Mon.\ Not.\ Roy.\ Astron.\ Soc.\  {\bf 425}, no. 2, 1597 (2012)]
  [arXiv:1103.0576 [astro-ph.CO]].

\bibitem{Pshirkov:2009vb}
  M.~S.~Pshirkov and A.~V.~Tuntsov,
  Phys.\ Rev.\ D {\bf 81}, 083519 (2010)
  [arXiv:0911.4955 [astro-ph.CO]].

\bibitem{Tuntsov:2010fu}
  A.~V.~Tuntsov and M.~S.~Pshirkov,
  Phys.\ Rev.\ D {\bf 81}, 063523 (2010)
  [arXiv:1001.4580 [astro-ph.CO]].

\bibitem{Olmez:2010bi}
  S.~Olmez, V.~Mandic and X.~Siemens,
  Phys.\ Rev.\ D {\bf 81}, 104028 (2010)
  [arXiv:1004.0890 [astro-ph.CO]].

\bibitem{Sanidas:2012ee}
  S.~A.~Sanidas, R.~A.~Battye and B.~W.~Stappers,
  Phys.\ Rev.\ D {\bf 85}, 122003 (2012)
  [arXiv:1201.2419 [astro-ph.CO]].

\bibitem{Binetruy:2012ze}
  P.~Binetruy, A.~Bohe, C.~Caprini and J.~F.~Dufaux,
  JCAP {\bf 1206}, 027 (2012)
  [arXiv:1201.0983 [gr-qc]].

\bibitem{Kogut:2011xw}
  A.~Kogut {\it et al.},
  JCAP {\bf 1107}, 025 (2011)
  [arXiv:1105.2044 [astro-ph.CO]].

\bibitem{Lorimer:2007qn}
  D.~R.~Lorimer, M.~Bailes, M.~A.~McLaughlin, D.~J.~Narkevic and F.~Crawford,
  Science {\bf 318}, no. 5851, 777 (2007)
  [arXiv:0709.4301 [astro-ph]].

\bibitem{Keane:2011mj}
  E.~F.~Keane, M.~Kramer, A.~G.~Lyne, B.~W.~Stappers and M.~A.~McLaughlin,
  Mon.\ Not.\ Roy.\ Astron.\ Soc.\  {\bf 415}, 3065 (2011)
  [arXiv:1104.2727 [astro-ph.SR]].

\bibitem{Thornton:2013iua}
  D.~Thornton {\it et al.},
  Science {\bf 341}, no. 6141, 53 (2013)
  [arXiv:1307.1628 [astro-ph.HE]].

\bibitem{Burke-Spolaor:2014rqa}
  S.~Burke-Spolaor and K.~W.~Bannister,
  Astrophys.\ J.\  {\bf 792}, no. 1, 19 (2014)
  [arXiv:1407.0400 [astro-ph.HE]].

\bibitem{Petroff:2014taa}
  E.~Petroff {\it et al.},
  Mon.\ Not.\ Roy.\ Astron.\ Soc.\  {\bf 447}, no. 1, 246 (2015)
  [arXiv:1412.0342 [astro-ph.HE]].

\bibitem{Ravi:2014mma}
  V.~Ravi, R.~M.~Shannon and A.~Jameson,
  Astrophys.\ J.\  {\bf 799}, no. 1, L5 (2015)
  [arXiv:1412.1599 [astro-ph.HE]].

\bibitem{Champion:2015pmj}
  D.~J.~Champion {\it et al.},
  Mon.\ Not.\ Roy.\ Astron.\ Soc.\  {\bf 460}, no. 1, L30 (2016)
  [arXiv:1511.07746 [astro-ph.HE]].

\bibitem{Keane:2016yyk}
  E.~F.~Keane {\it et al.},
  Nature {\bf 530}, 453 (2016)
  [arXiv:1602.07477 [astro-ph.HE]].

\bibitem{Ravi:2016kfj}
  V.~Ravi {\it et al.},
  Science {\bf 354}, no. 6317, 1249 (2016)
  [arXiv:1611.05758 [astro-ph.HE]].

\bibitem{Spitler:2014fla}
  L.~G.~Spitler {\it et al.},
  Astrophys.\ J.\  {\bf 790}, no. 2, 101 (2014)
  [arXiv:1404.2934 [astro-ph.HE]].

\bibitem{Masui:2015kmb}
  K.~Masui {\it et al.},
  Nature {\bf 528}, 523 (2015)
  [arXiv:1512.00529 [astro-ph.HE]].

\bibitem{Falcke:2013xpa}
  H.~Falcke and L.~Rezzolla,
  Astron.\ Astrophys.\  {\bf 562}, A137 (2014)
  [arXiv:1307.1409 [astro-ph.HE]].

\bibitem{Totani:2013lia}
  T.~Totani,
  Pub.\ Astron.\ Soc.\ Jpn.\  {\bf 65}, L12 (2013)
  [arXiv:1307.4985 [astro-ph.HE]].

\bibitem{Kashiyama:2013gza}
  K.~Kashiyama, K.~Ioka and P.~Meszaros,
  Astrophys.\ J.\  {\bf 776}, L39 (2013)
  [arXiv:1307.7708 [astro-ph.HE]].

\bibitem{Pen:2015ema}
  U.~L.~Pen and L.~Connor,
  Astrophys.\ J.\  {\bf 807}, no. 2, 179 (2015)
  [arXiv:1501.01341 [astro-ph.HE]].

\bibitem{Patterson:2008ie}
  C.~D.~Patterson, S.~W.~Ellingson, B.~S.~Martin, K.~Deshpande, J.~H.~Simonetti, M.~Kavic and S.~E.~Cutchin,
  ACM Trans.\ Reconf.\ Tech.\ Syst.\  {\bf 1}, 1 (2009)
  [arXiv:0812.1255 [astro-ph]].

\bibitem{Henning:2010aa}
  P.~Henning {\it et al.},
  PoS ISKAF {\bf 2010}, 024 (2010)
  [arXiv:1009.0666 [astro-ph.IM]].

\bibitem{Fender:2008zh}
  R.~Fender {\it et al.} [LOFAR Collaboration],
  PoS Dynamic {\bf }, 030 (2007)
  [arXiv:0805.4349 [astro-ph]].

\bibitem{Nan:2011um}
  R.~Nan {\it et al.},
  Int.\ J.\ Mod.\ Phys.\ D {\bf 20}, 989 (2011)
  [arXiv:1105.3794 [astro-ph.IM]].

\bibitem{Miyamoto:2012ck}
  K.~Miyamoto and K.~Nakayama,
  JCAP {\bf 1307}, 012 (2013)
  [arXiv:1212.6687 [astro-ph.CO]].

\bibitem{Steer:2010jk}
  D.~A.~Steer and T.~Vachaspati,
  Phys.\ Rev.\ D {\bf 83}, 043528 (2011)
  [arXiv:1012.1998 [hep-th]].

\bibitem{BlancoPillado:2000xy}
  J.~J.~Blanco-Pillado and K.~D.~Olum,
  Nucl.\ Phys.\ B {\bf 599}, 435 (2001)
  [astro-ph/0008297].

\bibitem{Bhat:2004xt}
  N.~D.~R.~Bhat, J.~M.~Cordes, F.~Camilo, D.~J.~Nice and D.~R.~Lorimer,
  Astrophys.\ J.\  {\bf 605}, 759 (2004)
  [astro-ph/0401067].

\bibitem{Lorimer:2013roa}
  D.~R.~Lorimer, A.~Karastergiou, M.~A.~McLaughlin and S.~Johnston,
  Mon.\ Not.\ Roy.\ Astron.\ Soc.\  {\bf 436}, 5 (2013)
  [arXiv:1307.1200 [astro-ph.HE]].

\bibitem{Caleb:2015uuk}
  M.~Caleb, C.~Flynn, M.~Bailes, E.~D.~Barr, R.~W.~Hunstead, E.~F.~Keane, V.~Ravi and W.~van Straten,
  Mon.\ Not.\ Roy.\ Astron.\ Soc.\  {\bf 458}, no. 1, 708 (2016)
  [arXiv:1512.02738 [astro-ph.HE]].

\bibitem{Ioka:2003fr}
  K.~Ioka,
  Astrophys.\ J.\  {\bf 598}, L79 (2003)
  [astro-ph/0309200].

\bibitem{Bera:2016qfe}
  A.~Bera, S.~Bhattacharyya, S.~Bharadwaj, N.~D.~R.~Bhat and J.~N.~Chengalur,
  Mon.\ Not.\ Roy.\ Astron.\ Soc.\  {\bf 457}, no. 3, 2530 (2016)
  [arXiv:1601.05410 [astro-ph.HE]].

\bibitem{Petroff:2016tcr}
  E.~Petroff {\it et al.},
  Publ.\ Astron.\ Soc.\ Austral.\  {\bf 33}, 45 (2016)
  [arXiv:1601.03547 [astro-ph.HE]].

\bibitem{Chatterjee:2017dqg}
  S.~Chatterjee {\it et al.},
  Nature {\bf 541}, 58 (2017)
  [arXiv:1701.01098 [astro-ph.HE]].

\bibitem{Brandenberger:2017uwo}
  R.~Brandenberger, B.~Cyr and A.~V.~Iyer,
  arXiv:1707.02397 [astro-ph.CO].

\end{thebibliography}
\end{document}